\documentclass{iau_FM}

\usepackage[T1]{fontenc}
\usepackage{amsmath}
\usepackage{graphicx}
\usepackage{multirow}

\title[Indigenous knowledges and kinship as a model for our future in outer space] 
{Indigenous knowledges and kinship as a model for our future in outer space}

\author[Hilding Neilson]   
{Hilding Neilson$^1$}

\affiliation{$^1$Department of Physics \& Physical Oceanography, \\ Memorial University of Newfoundland \& Labrador,
St.~John's, NL A1B 3X7, Canada\\ email: {\tt hneilson@mun.ca} \\[\affilskip]
}

\pubyear{2024}
\jname{Astronomy in Focus, Volume 1} 
\editors{Piero Benvenuti, ed.}

\begin{document}

\maketitle
\begin{abstract}
Commercial endeavours have already compromised our relationship with space. The Artemis Accords are creating a framework that will commercialize the Moon and further impact that relation. To confront that impact, a number of organizations have begun to develop new principles of sustainability in space, many of which are borne out of the capitalist and colonial frameworks that have harmed water, nature, peoples and more on Earth. Indigenous methodologies and ways of knowing offer different paths for living in relationship with space and the Moon. While Indigenous knowledges are not homogeneous, there are lessons we can use from some of common methods. In this talk we will review some Indigenous methodologies, including the concept of kinship and discuss how kinship can inform our actions both on Earth and in space.
\end{abstract}


\maketitle

\section{Introduction}
The Outer Space Treaty (\footnote{https://www.unoosa.org/oosa/en/ourwork/spacelaw/treaties/outerspacetreaty.html}) notes that activities in outer space should be for the ``benefit of all mankind''.  However, the current explosive growth of the private space industry has impacted the human relationship with the night sky, celestial objects such as the Moon, and outer space itself.  In that sense, not all of humanity seems to benefit from these endeavours. In particular, a number of actors, including the International Astronomical Union, have been developing new reports and conferences with the aim of mitigating the impact of a minimally regulated space industry including the SATCON1 and SATCON2 reports led by the National Science Foundation's National Optical-Infrared Astronomy Research Laboratory (\cite[Walker et al. 2020b, Hall et al. 2021]{Sat1, Sat2})  and the Dark \& Quiet Skies for Science and Society conferences led by the International Astronomical Union (IAU) and the United Nations Office of Outer Space Affairs and others (\cite[Walker et al. 2020a]{DQS1}).  In these reports, Indigenous peoples are treated as a stakeholder along with planetaria and amateur astronomy groups and in some of the reports as victims of this growing light and space pollution. In the latter sense arguments are made that light pollution will negatively impact ancient practices and connections with the night sky, usually without input from Indigenous peoples and ignoring that Indigenous peoples have modern cultures. 

In this presentation, I aim to consider the discussion of protecting the night sky and outer space through a lens of  Indigenous methodologies as a third perspective from the current binary discussion that pits outer space as only an exploitable resource that benefits the growing space industrial complex against the argument led by astronomers that outer space must be treated as a natural  environment that must be protected and be untouched (e.g., \cite[Lawrence et al. 2022]{lawrence2022case}).  This third method is motivated from the perspective that both sides of current discussion are colonial, and that both minimize the rights of Indigenous peoples with respect to outer space.  To that end this presentation is designed to advocate for {\bf policy changes} that centre and include Indigenous peoples, nations, and methods and moves beyond community engagement (\cite[Barentine \& Heim 2023]{Barentine2023}).

Before continuing the discussion, it is necessary to acknowledge that the author is Mi'kmaq and Settler and can only speak to that experience.  Furthermore, as a professional astronomer the author acknowledges that the field of professional astronomy has benefited from past, current, and ongoing colonization, particularly in Chile \footnote{https://astrobites.org/2019/09/10/astronomical-observatories-and-indigenous-communities-in-chile/}, Hawai'i (\cite[Salazar 2014]{Salazar2014}), and elsewhere.  There is an irony that the professional astronomy community seeks to engage with Indigenous peoples with respect to activities in outer space while ignoring and dismissing Indigenous peoples when building observatories.

\section{The coloniality of space exploitation and space environmentalism}
Much of the current discourse in the current era of outer space exploration that has become a narrative of space exploitation.  This narrative has been growing with the Artemis Accords \footnote{https://www.nasa.gov/artemis-accords/} that seek to provide a set of guidelines for resource extraction in outer space and on the Moon, with an eye to asteroid mining and Martian exploration. The Accords, while designed to not directly conflict with the Outer Space Treaty, allows for operators to remove material from bodies in outer space and not to be interfered with in locations where operators have equipment and satellites.  The Accords claim that no private operator or nation state can claim territory in outer space, yet they allow for operators to remove material and effectively 'squat'. Even if this does not mean ownership in outer space, it follows a path of the 19th-century Gold Rush where prospectors staked claims and took whatever was considered valuable while often leaving nothing but waste. In that case, many of these gold rushes significantly harmed Indigenous peoples, stole Indigenous lands, and had significant negative impact on the local plants, animals, and waters that had been there since time immemorial.
As such the Artemis Accords are colonial.

That colonialism is not new.  \cite[Trevi\~no (2020)]{Trevino} noted how the history of space exploration is part of a colonial narrative that has existed for centuries.  From the first manual on the colonization of Mars by Werner Von Braun, the development of outer space activities have been justified by the same narrative that colonists used to settle the Americas: terra nullius, and Manifest Destiny.  In many ways the growth of light pollution around the world is also part of this colonialism and terra nullius.  \cite[Hamacher et al. (2020)]{Hamacher} wrote that "Light Pollution is Cultural Genocide" on the principle that many Indigenous nations and peoples write stories in their constellations (both star-based and ``dark'' constellations). These stories carry significant cultural, scientific and holistic knowledges that continue to be a crucial part of those cultures.  In the same way that light pollution is colonisation, so too is light pollution from satellites.  Whiles light pollution acts to erases Indigenous knowledges and stories, the growth of bright satellites aim to rewrite them.  This, too, is a significant colonial problem.

On the other hand, an increasingly commonly touted solution for regulation is to treat outer space as part of the environment.  While this might appear to be a way to include Indigenous peoples and methods in the conversation about regulating outer space, much of the environmental movement has been colonial. One such example is the development of the National Park systems in Canada and the United States that stole land from Indigenous peoples in the name of conservationism.  It is notable that the clause in the Artemis Accords citing the need to preserve sites of cultural importance will potentially act in the same manner, creating ``National Parks'' on the Moon based on the interests of Eurocentric cultures. 

Furthermore, how this outer space environmentalism will be implemented will arguably be centred on the interests of nation states, especially the richest states that can afford the investment necessary for outer space exploitation.  Part of the space interests of these states is professional astronomy, and hence will include the need to support ground-based astronomy.  Since most of the largest radio and optical telescopes that exist or are planned are on Indigenous lands, then any outer space environmental movement could easily have negative consequences for Indigenous peoples.

\section{Outer space is part of the land}
Indigenous peoples have been connected to their homelands since time immemorial.  That connection includes a deep, holistic relationship with the night sky, the motions of the Moon and the planets and more.  That connection includes oral traditions that use the night sky for rituals, calendars, ethics, and more. Many peoples tell stories of the North Star as a guide star, and many in the Pacific use the stars for way-finding (\cite[Ruggles  2015]{Ruggles2015}).  In the far north,  Inuit peoples use the Flickering Star as a guide for weather prediction (\cite[MacDonald \& Schledermann  1998]{Macdonald}), while in the southern hemisphere, many nations use dark constellations for time-keeping (\cite[Urton 1981]{Urton}).  The Mi'kmaw people use the constellation of Muin and the Seven Bird Hunters to share about circumpolar stars, hunting ethics, biology and more (\cite[Marschall et al. 2010]{Bartlett}). The night sky is an important part of Indigenous knowledges and cultures since time immemorial and  for many peoples the night sky is a reflection of the land below.

Space exploration is also part of the oral traditions.  For instance, many stories about the constellation that is called the Pleiades in the Eurocentric tradition is about travellers to the sky in a number of Indigenous traditions (\cite[Lankford 2007]{lankford2007reachable}).  In many cultures, the people come from the sky (\cite[Buck  2018]{Buck}). Indigenous oral traditions contain many stories and knowledges about people who explore the Moon and outer space.  That exploration does not occur in a physical sense that we tend to consider space exploration, i.e., foot- and rover prints on celestial bodies, nor does it occur in the same sense that academic astronomy explores outer space through the interpretation of photons and gravitational waves.  This space exploration occurs in ways centred on Indigenous methodologies that must also be part of any conversation about outer space exploration and exploitation if we are to act in anti-colonial ways.

These relationships denote deep connections with outer space for navigation, time-keeping, and more.  For many Indigenous nations, these connections come with protocols, treaties, and responsibilities. \cite{SkyCountry} note that for Yol{\it{n}}u peoples in Australia, interactions with Sky Country requires people to follow protocols.  In Mi'kmaw cosmology of six worlds, the sky above is its own world (\cite[Whitehead \& Kaubach 1988]{Whitehead}). As such, interactions between L'nu\footnote{L'nu is the Mi'kmaq word for people.} and the sky above is governed by treaties, similar to treaties between different Indigenous nations or with different animals, plants, water, and more. These treaties outline both rights and responsibilities for all sides.  In that situation, honouring those responsibilities would preclude peoples from the pure exploitation of outer space, but would not necessarily require ending any form of space exploration in the Eurocentric sense.

The combination of interactions with outer space since time immemorial along with the protocols and treaties in outer space mean that Indigenous peoples and nations have sovereignties and responsibilities with respect to the night sky and how we interact with outer space.  {\bf Therefore, however nation states choose to agree to exploit outer space, there must be consideration for Indigenous rights and methodologies.}

\section{Indigenous methodologies in outer space}
Much of how we view our place with respect to exploring and exploiting outer space derives from the Western perspective of sciences and the methods used.  However, Indigenous methodologies offer different perspectives on the human relationship with outer space that would be a third way between the current perspectives of outer space Conservation or outer space exploitation. 

The concept of Indigenous methods or sciences have been discussed by numerous authors in many different research fields (e.g., \cite[Cajete 2000; Battiste 2013; and Lipe 2019]{Cajete2000, battiste2013decolonizing, Lipe2019}).  To summarize, one can list these Indigenous methods of knowing to be:\\
\begin{itemize}
\item What's above reflects below. \\
This concept is a comparison of how humans are related to what happens in the sky.  For instance, many constellations reflect stories of nature and the land, be it the story of Muin and the Seven Bird Hunters in Mi'kmak'i (\cite[Marshall et al. 2010]{Bartlett})  and the Cree sky stories (\cite[Buck 2018]{Buck}).  These stories speak to relationships between the land and the night sky and how perspectives depend on where on the land one is.\\
\item Knowledge is relational. \\
For many Indigenous peoples, knowledge is not something universal but depends on the person learning and their perspectives.  Furthermore, knowledge is relational because knowledge is something that is used; knowledge is active and to be shared ethically.  Knowledge is not something that is simply obtained.\\
\item Indigenous methods considers multiple variables that interact concurrently.\\
In traditional Western/Eurocentric sciences, the method to understand a phenomenon is to attempt to reduce it to the minimal number of variables. To reduce variables to the minimal number  usually requires measuring average values and missing how some of those variables interact.  By considering multiple  variables concurrently, knowledge keepers and Elders develop and grow the ability to predict and manipulate phenomena. \\
\item Knowledge is (w)holistic.\\
For many Indigenous peoples knowledges have many purposes. A star story is more than what might be classified solely as astronomy.  Star stories such as Muin and the Seven Bird hunters reflect biology and nature, ethics and community along the motions of stars.  As such Indigenous knowledges tend to not be placed into disciplines akin to how traditional Western/Eurocentric knowledges are.\\
\item Nature  is holy, sacred, and familial. \\
In traditional Western sciences, nature is usually treated as a hierarchy. In that sense, we tend to place humans and human rights above the rights of animals, plants, rocks, etc.  One way this is seen is how elements of nature are treated as resources for the service of humans.  For many Indigenous nations, nature is animate and has rights.  Interactions with nature operate in reciprocity where people give back when they take from nature (\cite[Kimmerer 2013]{Braiding}).  Another way to view this is through the lens of treaty.  In a treaty with nature, Indigenous peoples have responsibilities to support nature  and the relationship is give and take.  In the Eurocentric model the relationship is to take and control.\\
\end{itemize}

It must be noted that there is no such thing as a pan-Indigenous knowledge system, there are Mi'kmaw ways of knowing, Salish ways of knowing and so on.  Every culture and nation will have its own methodologies  and relationships with nature.  But, there are commonalities that can be considered such as the methods listed.  These methods are significantly different Eurocentric/Western methods in how humans relate with nature, relate with the sky, and, most importantly, our responsibilities we owe nature and the sky. As such, many Indigenous peoples consider the stars, planets and the Moon to be a relation, or an ancestor.  This perspective is different from the Western perspective where the Moon is a dead rock, the stars are not alive and so on. 

As such, these methods suggest a different way to interact and live with the night sky, the planets, and the Moon, and how we would operate in outer space.  Since the planets and the Moon are relations or kin then these bodies have rights and are alive or animate in a sense that is antithetical to Western/Eurocentric methods.  Furthermore, this also suggests that the outer space is a land onto itself that requires relationship building and  can be claimed or abused by humans.  This is completely different from the two current leading narratives about humanity's future in outer space. 

One narrative is the traditional colonial narrative that it is humanity's destiny to travel into outer space and settle other worlds: a Manifest Destiny of outer space.  That Manifest Destiny is being accelerated by the Artemis Accords that invite private interests to mine and use outer space, treating it as a resource to be exploited and to profit from.  The second narrative is outer space as an environment that must be protected for sustainable usage by humans.  This narrative seeks to create protections but environmentalism on Earth can be colonial and anti-Indigenous.  The primary example of this the development of national parks in Canada and the United States that removed Indigenous peoples from their traditional lands to protect those lands for the common good.  While there are additional examples that can be cited, the key issue regarding both of these narratives is that they both claim to seek to use outer space, the Moon and more for the benefit of humanity.  

The Indigenous ways of knowing cited here tell us that there is a third way to proceed where we seek to act in treaty or relationship with outer space as an land where humans are guests when we operate there.  This act of treaty does not mean humanity must cease any operations in outer space, but that those operations must be carried out as good guests.  This may seem like a strange concept, but it is consistent with how Indigenous peoples lived on the lands of Turtle Island (North America) and beyond for millennia (\cite[Kimmerer 2013]{Braiding}). 

What would being a good guest in outer space look like?  It is not obvious, but operators in outer space would not simply take and satellite operators would not simply fill the orbital space around the Earth with satellite constellations, and it would require minimising the existence of `dead' satellites in orbit.  Being a good guest requires operators and users in outer space to given back and support the nature of outer space, the Moon and celestial objects. How we give back to outer space is something that humanity and nations can work on developing beyond debating who decides how outer space will be used to serve humanity or, more likely, serve a privileged few.

\section{Summary}
The goal of this work is to illustrate a third path to how humans can approach our relationships with and actions in outer space based on Indigenous methods and ways of knowing.  In this approach, I suggest that one way to be more consistent with these Indigenous methods to consider our treaties and relationships as seeking to not treat outer space as a resource for the benefit or profit of humanity, but, instead, that we live in treaty with outer space as its own land and place with its own rights of existence.  While, this behaviour would be dramatically different
than what humanity has done previously in outer space and is doing today, it is possible to change our systems and allow humans and Western societies to live in relationship with outer space.  I argue that the first step in building this relationship is to develop protocols for what humans can give to outer space when we take from it, whether this taking is of water from the Moon, minerals from asteroids or even from the near-vacuum of outer space when we place satellites.

While this work illustrates how Indigenous methods can inform policy regarding outer space, but more than that it notes how Indigenous peoples and nations should be a part of any policy with respect to the usage of outer space and that participation should be equitable and respectful.


\end{document}